# Predictability of climate anomalies in the regions of Northern Eurasia in the spring-summer months in 2024 in connection with El Niño

Mokhov I.I.
A.M. Obukhov Institute of Atmospheric Physics RAS
Lomonosov Moscow State University
mokhov@ifaran.ru

**Abstract**

The predictability of climate anomalies in the regions of Northern Eurasia in connection with El Niño phenomena is analyzed. Particular attention is paid to the most likely transition in 2024 from an El Niño phase at the beginning of the year to a La Niña phase at the end of the year, with the greatest probability of high temperatures and dry conditions in European Russia during the spring and summer months, as in 2010. The predictability levels of regional climate anomalies using different El Niño indices are compared. The relationship of the noted seasonal anomalies with atmospheric blockings is considered, taking into account the different phases of the key modes of climate variability like El Niño phenomena and the Pacific Decadal Oscillation. Changes in the predictability of regional climate anomalies under global climate change are discussed.

**Introduction**

Possible anomalies of surface air temperature, precipitation and drought conditions in the Russian regions in the spring-summer months of 2024 are estimated similarly to [1–5] using long-term regional seasonal data and taking into account the El Niño phase (*E*-phase) at the beginning of the year.

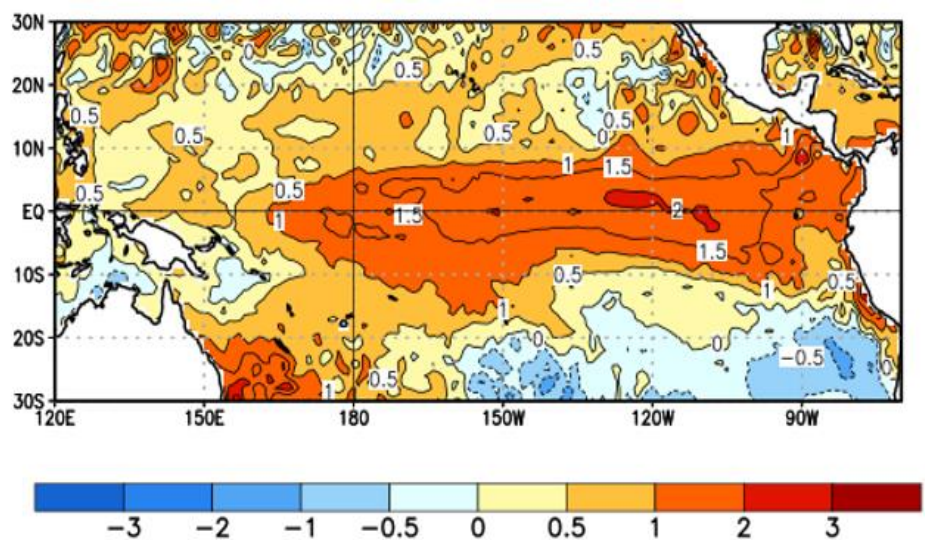

Fig. 1. Average SST anomalies (K) in the equatorial Pacific Ocean during 21/01/2024-17/02/2024 [6].

Figure 1 shows sea surface temperature (SST) anomalies (K) in the equatorial Pacific Ocean for the period 21/01/2024-17/02/2024 [6].

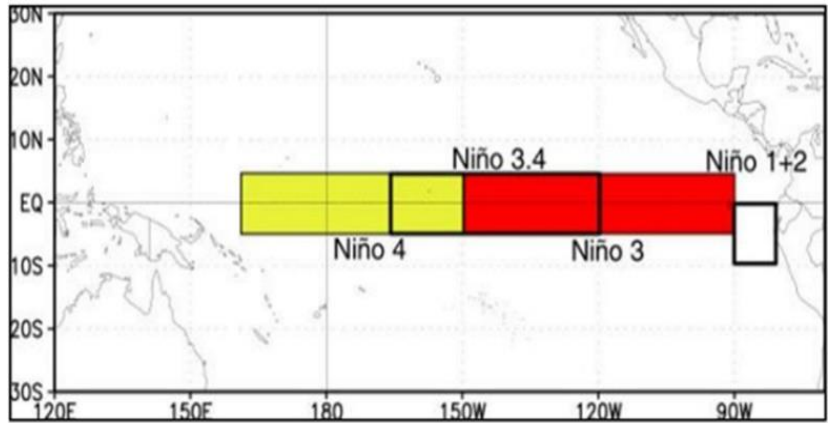

Fig. 2. Niño3, Niño4 and Niño3.4 regions [6].

To assess the effects of El Niño, their indices were used, characterized by the sea surface temperature in the regions of Niño3, Niño3.4 and Niño4 in the equatorial latitudes of the Pacific Ocean (Fig. 2).

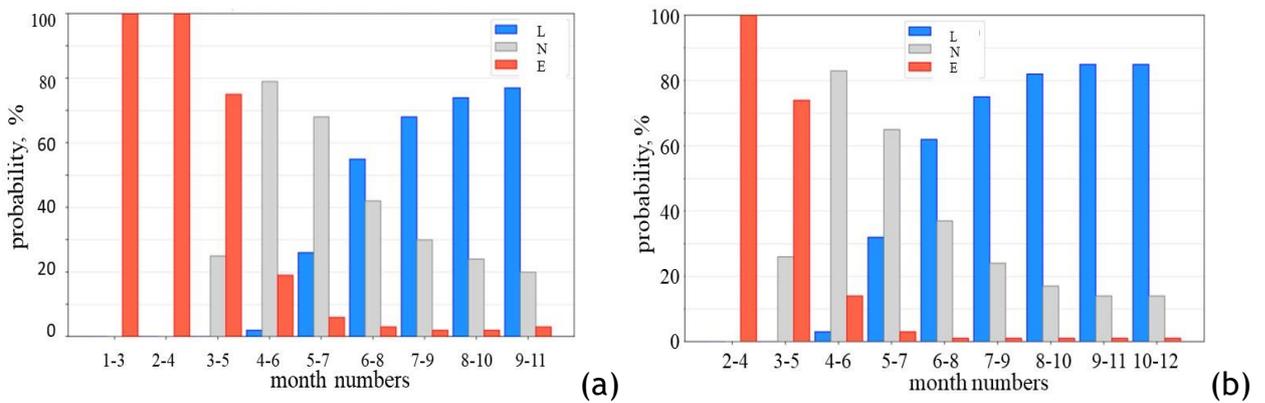

Fig. 3. Forecast model calculations for El Nino event: CPC probabilistic ENSO outlook updated on 08 February 2024 (a) and 14 March 2024 (b) [6].

Also, the results of forecast model calculations made in February and March 2024 [6] were taken into account (Fig. 3,4). Figure 3 shows results of model calculations with a transition from El Niño to ENSO-neutral is expected by April-June season 2024, with ENSO-neutral persisting through May-July 2024 presented in [6]. Thereafter, La Niña is favored in June-August, and chances increase through the September-November season.

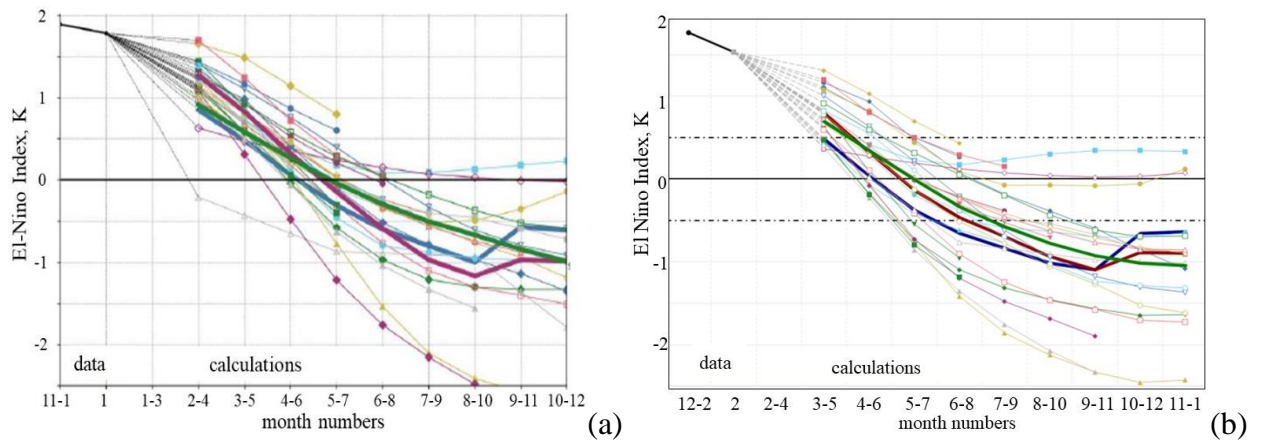

Fig. 4. Ensemble statistical and dynamical model calculations for the El Niño event forecast: IRI Pacific Nino3.4 SST model outlook updated on 19 February 2024 (a) and 19 March 2024 (b) [6].

According to ensemble model calculations on Fig. 3, the probability of transition to the *L*-phase by the end of 2024 (*E*→*L* transition) is expected to be about 80% to the beginning of March 2024 and about 90% to the end of March 2024. The corresponding probability for the neutral *N*-phase (*E*→*N* transition) is estimated to be less than 20% to the beginning of March 2024 and about 10% to the end of March 2024. The probability for the *E*-phase (*E*→*E* transition) is even lower.

Figure 4 shows results of ensemble statistical and dynamical model calculations: IRI Pacific Nino3.4 SST model outlook updated on 19 February 2024 (a) and 19 March 2024 (b) [6]. The majority of models indicate El Niño will persist through March-May 2024 and then transition to ENSO-neutral during April-June 2024. After a brief period of ENSO-neutral conditions, most models indicate a transition to La Niña around June-August 2024.

**Data and method used**

The spring and summer anomalies of surface air temperature $\delta T$, precipitation $dP$, drought index $D$, and excessive moisture index $M$ were analyzed for the European (ER) and Asian (AR) mid-latitudes for 1891–2015 [7,8] (see also [1-5]). According to [7,8], the drought index $D$ for the mid-latitude regions of Russia characterizes the portion of the area with positive anomalies of temperature of not less than 1 K and with negative anomalies of precipitation of not less than 20%; the index $M$ characterizes the portion of the area of ER and AR with negative anomalies of temperature of not less than 1 K and with positive anomalies of precipitation of not less than 20%.

Due to the fact that different types of El Niño are manifested, it is required to analyze the degree of agreement or disagreement between the estimates of probability of regional climate anomalies depending on the El Niño index. In particular, along with the canonical manifestations of El Niño with significant positive anomalies of SST in the eastern equatorial Pacific (EP El Niño), El Niño events with the most significant anomalies of SST in the central equatorial Pacific (CP El Niño) have often been observed in recent decades [9,10]. To assess the El Niño (La Niña) effects, the indices characterized by SST in the region Niño3 (90°–150° W) for EP El Niño and in the region Niño4 (160° E–150° W) in the equatorial Pacific for CP El Niño were used (ftp://www.coaps.fsu.edu/pub/).

The El Niño phase (*E*) and the opposite phase (with negative SST anomalies) La Niña (*L*) were identified using the 5-month moving averaging of SST anomaly values. The El Niño phase (warm phase) and La Niña phase (cold phase) were defined by the values of SST anomalies of not less than 0.5°C or not more than –0.5°C during six consecutive months. The other cases were identified as a neutral phase (*N*).

Similar to [1,2], nine possible transition regimes can be separated. For the years starting in the phase *N*, three transitions are possible during a year: *N*→*E* characterizes the transition from the phase *N* at the beginning of the year to the phase *E* at the beginning of the next year; *N*→*L* characterizes the respective transition from the phase *N* to the phase *L*; *N*→*N* means the continuation of the phase *N*. For the years starting in the phase *E*, three transitions are possible during the year: *E*→*E*, *E*→*L*, and *E* → *N*; the transitions *L*→*E*, *L*→*L*, and *L*→*N* are possible for the years starting in the phase *L*.

If Niño3 and Niño4 indices are used for the classification of transitions, the number of years starting with the phase *N* for the period of 1891–2015 (125 years) is equal to 68. The years starting with the phase *E* and *L* make up less than a half of all analyzed years: 28 or 29 depending on the index used (Niño3 or Niño4). The most frequent transitions are *N*→*N*, and the least frequent ones are *E*→*E* (in case of Niño3) and *L*→*E* (in case of Niño4): the average frequency is more than 30 years [3,4].

To assess the risk of possible regional climate anomalies, the results of ensemble model forecasts of El Niño evolution during a year [6] were used along with the observational data [3,4].

**Results**

Table 1 shows estimates of the probability of temperature anomalies δ*T* in ER and AR in May-June-July (MJJ) for various transitions from the *E*-phase at the beginning of the year using different El Niño indices. According to these estimates, for the *E*→*L* transition positive temperature anomalies in the ER were remarkably more frequent than negative temperature anomalies. Also, positive temperature anomalies in the ER are more probable than in the AR. Extreme positive temperature anomalies in the ER are several times more probable than in the AR for the *E*→*L* transition [3].

Table 1. Probability of positive and negative surface air temperature anomalies (δ*T*) in the ER (and AR) in MJJ for different transitions from La-Nina conditions at the beginning of the year (characterized by indices Nino3 and Nino4) from observations since 1891 (n – number of years).

| δ*T*, K ER (AR) | | >0 | | | | ≤0 | | | |
|---|---|---|---|---|---|---|---|---|---|
| | | > 0 | | > 1 K | | ≤ 0 | | < -1 K | |
| Nino3 *n*=28 | *E*→*E* *n*=4 | 17/28 (12/28) | **3/4** (2/4) | 8/28 (7/28) | 1/4 (2/4) | 11/28 (**16/28**) | 1/4 (2/4) | 5/28 (7/28) | 0/4 (1/4) |
| | *E*→*N* *n*=15 | | 6/15 (6/15) | | 2/15 (4/15) | | **9/15** (**9/15**) | | 5/15 (4/15) |
| | *E*→*L* *n*=9 | | **8/9** (4/9) | | 5/9 (1/9) | | 1/9 (**5/9**) | | 0/9 (2/9) |
| Nino3.4 *n*=36 | *E*→*E* *n*=9 | 21/36 (**20/36**) | **5/9** (**5/9**) | 9/36 (9/36) | 1/9 (4/9) | 15/36 (16/36) | 4/9 (4/9) | 7/36 (7/36) | 3/9 (3/9) |
| | *E*→*N* *n*=15 | | 6/15 (7/15) | | 3/15 (4/15) | | **9/15** (**8/15**) | | 4/15 (2/15) |
| | *E*→*L* *n*=12 | | **10/12** (8/12) | | 5/12 (1/12) | | 2/12 (4/12) | | 0/12 (2/12) |
| Nino4 *n*=29 | *E*→*E* *n*=8 | 17/29 (**16/29**) | 3/8 (**6/8**) | 10/29 (9/29) | 2/8 (**6/8**) | 12/29 (13/29) | **5/8** (2/8) | 7/29 (5/29) | 3/8 (1/8) |
| | *E*→*N* *n*=13 | | 7/13 (5/13) | | 4/13 (2/13) | | 6/13 (**8/13**) | | 4/13 (3/13) |
| | *E*→*L* *n*=8 | | **7/8** (**5/8**) | | 4/8 (1/8) | | 1/8 (3/8) | | 0/8 (1/8) |

Corresponding estimates of the probability of precipitation anomalies δP in ER and AR in MJJ for various transitions from the E-phase at the beginning of the year using various El Niño indices are presented in Table 2. According to Table2, for the most probable transition E→L in 2024, estimates of the probability of positive and negative precipitation anomalies in the ER and AR differ less significantly than for temperature anomalies.

Table 2. Probability of positive and negative surface air temperature anomalies (δ*P*) in the ER (AR) in MJJ for different transitions from La-Nina conditions at the beginning of the year (characterized by indices Nino3 and Nino4) from observations since 1891 (n – number of years).

| δP, % | | <0 | | | ≥0 | | |
|---|---|---|---|---|---|---|---|
| ER (AR) | | <0 | | < -20% | ≥0 | | >20% |
| Nino3 n=28 | E→E n=4 | 9/28 (**16/28**) | 0/4 (2/4) | 4/28 (4/28) | 0/4 (0/4) | 19/28 (12/28) | **4/4** (2/4) | 1/28 (3/28) | 0/4 (0/4) |
| | E→N n=15 | | 5/15 (**10/15**) | | 3/15 (2/15) | | **10/15** (5/15) | | 1/15 (0/15) |
| | E→L n=9 | | 4/9 (4/9) | | 1/9 (0/9) | | **5/9** (**5/9**) | | 0/9 (3/9) |
| Nino3.4 n=36 | E→E n=9 | 13/36 (**20/36**) | 2/9 (4/9) | 5/36 (8/36) | 1/9 (4/9) | 23/36 (16/36) | **7/9** (**5/9**) | 3/36 (4/36) | 0/9 (0/9) |
| | E→N n=15 | | 6/15 (**10/15**) | | 3/15 (2/15) | | 9/15 (5/15) | | 2/15 (0/15) |
| | E→L n=12 | | 5/12 (6/12) | | 1/12 (2/12) | | **7/12** (6/12) | | 1/12 (4/12) |
| Nino4 n=29 | E→E n=8 | 10/29 (**17/29**) | 1/8 (4/8) | 3/29 (4/29) | 0/8 (1/8) | 19/29 (12/29) | **7/8** (4/8) | 2/29 (4/29) | 0/8 (0/8) |
| | E→N n=13 | | 6/13 (**9/13**) | | 2/13 (1/13) | | 7/13 (4/13) | | 1/13 (1/13) |
| | E→L n=8 | | 3/8 (4/8) | | 1/8 (2/8) | | **5/8** (4/8) | | 1/8 (3/8) |

Table 3 shows estimates of the probability for different values of drought (D) and excessive moisture (M) indices in ER and AR in MJJ for various transitions from the *E*-phase at the beginning of the year using different El Niño indices [3].

Table 3. Estimates of probability of D and M in the ER (and AR) in MJJ for different transitions from La-Nina conditions at the beginning of the year (characterized by indices Nino3 and Nino4) from observations since 1891 (n – number of years).

| D, % | | D | | | | M | | | |
|---|---|---|---|---|---|---|---|---|---|
| ER (AR) | | ≥20% | | ≥30% | | ≥20% | | ≥30% | |
| Nino3 n=28 | E→E n=4 | **14/28** (11/28) | 3/4 (2/4) | **7/28** (5/28) | 0/4 (0/4) | 6/28 (6/28) | 0/4 (2/4) | **3/28** (2/28) | 0/4 (1/4) |
| | E→N n=15 | | 5/15 (6/15) | | 3/15 (4/15) | | 6/15 (2/15) | | 3/15 (0/15) |
| | E→L n=9 | | **6/9** (3/9) | | **4/9** (1/9) | | 0/9 (2/9) | | 0/9 (1/9) |
| Nino3.4 n=36 | E→E n=9 | **17/36** (13/36) | 4/9 (4/9) | **8/36** (6/36) | 0/9 (1/9) | 9/36 (6/36) | 3/9 (3/9) | **4/36** (2/36) | 2/9 (1/9) |
| | E→N n=15 | | 6/15 (5/15) | | 4/15 (4/15) | | 5/15 (1/15) | | 2/15 (0/15) |
| | E→L n=12 | | **7/12** (4/12) | | **4/12** (1/12) | | 1/12 (2/12) | | 0/12 (1/12) |
| Nino4 n=29 | E→E n=8 | 14/29 (14/29) | **4/8** (6/8) | 7/29 (7/29) | 0/8 (3/8) | **9/29** (5/29) | 4/8 (2/8) | **3/29** (2/29) | 1/8 (0/8) |
| | E→N n=13 | | 6/13 (5/13) | | 4/13 (3/13) | | 4/13 (2/13) | | 2/13 (2/13) |
| | E→L n=8 | | **4/8** (3/8) | | **3/8** (1/8) | | 1/8 (1/8) | | 0/8 (0/8) |

According to Table 3 in ER in MJJ during the *E→L* transition, the probability of a drought conditions with index D≥20% is estimated to be maximum (2/3) when El Niño phenomena are detected by the Nino3 index, and minimal (1/2) when detected by the Nino4 index. The corresponding probability of a drought conditions with index D≥30% is also estimated to be maximum (> 40%) when detecting El Niño phenomena with the Nino3 index, and minimal (1/3) when detected by the Nino3.4 index.

Table 4. Estimates of the frequency of positive and extreme positive (> 1K) SAT anomalies δT in MJJ using the indices Niño3, Niño3.4, and Niño4 for different periods: 1891–2015, 1950–2015 and 1980–2015 for ER during the $E{\rightarrow}L$ transition. The number in the denominator is the total number of corresponding phase transitions $E{\rightarrow}L$ for ER.

| ER        | Niño4 |       | Niño3.4 |       | Niño3 |       |
|-----------|-------|-------|---------|-------|-------|-------|
| $E{\rightarrow}L$ | > 0   | > 1K  | > 0     | > 1K  | > 0   | > 1K  |
| 1891–2015 | 7/8   | 4/8   | 10/12   | 5/12  | 8/9   | 5/9   |
| 1950–2015 | 7/7   | 4/7   | 8/9     | 5/9   | 6/7   | 4/7   |
| 1980–2015 | 5/5   | 4/5   | 6/6     | 5/6   | 4/4   | 4/4   |

A more detailed analysis was carried out for the transition $E{\rightarrow}L$, for which, for the total period 1891-2015, the highest estimates of the frequency of extreme values of surface air temperature in ER was revealed [4]. Table 4 presents estimates of the frequency of positive and extreme positive (>1 K) SAT anomalies δT in MJJ for ER during the $E{\rightarrow}L$ transition using the indices Niño3, Niño3.4 and Niño4 for different periods: 1891–2015, 1950–2015 and 1980–2015. The According to Table 1 the number of corresponding (> 0 or > 1 K) anomalies (in the numerator) and the total number of phase transitions (in the denominator) $E{\rightarrow}L$ for ER.

According to Table 4 the frequency of positive and extreme positive (>1 K) SAT anomalies δT in MJJ for ER during the $E{\rightarrow}L$ transition is increasing for the last decades (up to 100%).

**Discussion and conclusion**

The predictability of climate anomalies in the regions of Northern Eurasia in connection with El Niño phenomena was analyzed. Particular attention was paid to the most likely $E{\rightarrow}L$ transition in 2024 with the greatest probability of high temperatures and dry conditions in European Russia during the spring and summer months, similar to summer of 2010.

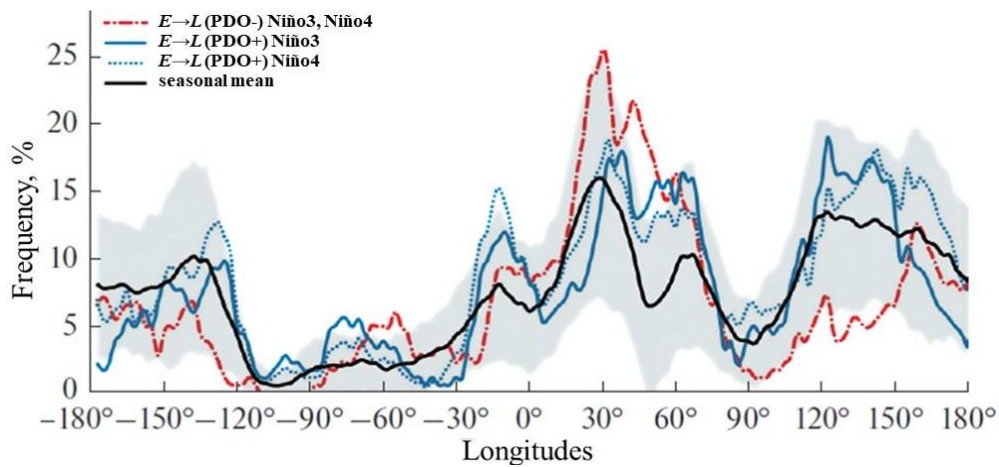

Fig. 5. Frequency of summer atmospheric blockings versus longitude in the Northern Hemisphere according to data for 1979–2019 for the $E{\rightarrow}L$ phase transitions detected by the Niño3 (solid curves) and Niño4 (dotted indices) indices during positive and negative PDO phases. The black curves characterize the average values of the summer blocking frequency over 1979–2019 (the standard deviation range is shaded).

The formation of noted seasonal anomalies is facilitated by atmospheric blockings against the background of a tendency for a regional decrease in precipitation, which accompanies an increase of surface air temperature [5]. Estimates of the blocking frequency $p$ for the $E{\rightarrow}L$ phase transitions by the Niño3 and Niño4 indices during the positive and negative PDO (Pacific Decadal Oscillation) phases are shown in Fig. 5 versus the average frequencies of summer atmospheric blockings versus the longitude in the midlatitudes of the Northern Hemisphere in 1979–2019. The $p$ anomalies significantly differ both when using different El Niño indices and in different PDO phases [11]. According to Fig. 6, $p$ values are maximal in the $E{\rightarrow}L$ transition (with an excess of the average longterm values by more than a standard deviation) over eastern Europe when both El Niño indices are used during the negative PDO phase (see, for instance: https://www.ncei.noaa.gov/pub/data/cmb/ersst/v5/index/ersst.v5.pdo.dat). Since 1979, two such transitions have been detected, in 2007 and 2010. Similar $E{\rightarrow}L$ phase transition in the negative PDO phase is expected in 2024. The record heat and fires in European Russia in summer 2010 were associated with the record long (about two months) atmospheric blocking conditions [12,13].

The predictability of noted regional weather and climate anomalies related with atmospheric blocking activity and key modes of climate variability (like El Niño and PDO) should depend on global climate changes. According to ensemble model estimates under warming in the 21st century an increase in the frequency of atmospheric blockings in the regions of the Northern Hemisphere is expected [14]. Also, significant differences in trends have been revealed for different types of El Nino phenomena, characterized by surface temperature anomalies in the equatorial latitudes of the eastern (Niño3) and central (Nino4) regions of the Pacific Ocean. In particular, the tendency of an increase in the frequency of occurrence of El Nino phases using a Niño4 index over past decades was noted. [15].